\documentclass[twocolumn,pra,aps,10pt,superscriptaddress]{revtex4-1}

\usepackage{graphicx}
\usepackage{amsmath}
\usepackage{amsfonts}
\usepackage{amssymb}
\usepackage{xcolor}
\usepackage{hyperref}
\hypersetup{colorlinks=true,allcolors=blue}
\usepackage[T1]{fontenc}

\usepackage{footnote}
\usepackage{textcomp}

\DeclareUnicodeCharacter{2009}{\,}

\newcommand{\mbf}[1]{\mathbf{#1}}
\renewcommand{\t}[1]{\textrm{#1}}
\newcommand{\nn}{\nonumber\\}

\newcommand{\q}{\mbf{q}}

\renewcommand{\a}{\alpha}
\renewcommand{\b}{\beta}
\newcommand{\g}{\gamma}

\renewcommand{\r}{\rho}
\newcommand{\s}{\sigma}

\newcommand{\w}{\omega}

\renewcommand{\L}{\mathcal{L}}

\newcommand{\+}{^\dagger}
\renewcommand{\>}{\rangle}
\newcommand{\<}{\langle}
\newcommand{\Tr}{\t{Tr}}

\newcommand{\G}{\vert G\>}
\newcommand{\X}{\vert X\>}

\newcommand{\XX}{\vert X\>\<X\vert}

\newcommand{\GX}{\vert G\>\<X\vert}

\begin{document}

\title{Schr\"{o}dinger cats in quantum-dot--cavity systems} 


\author{M. Cosacchi}
\email{michael.cosacchi@uni-bayreuth.de}
\affiliation{Theoretische Physik III, Universit{\"a}t Bayreuth, 95440 Bayreuth, Germany}
\author{T. Seidelmann}
\affiliation{Theoretische Physik III, Universit{\"a}t Bayreuth, 95440 Bayreuth, Germany}
\author{J. Wiercinski}
\affiliation{Theoretische Physik III, Universit{\"a}t Bayreuth, 95440 Bayreuth, Germany}
\author{M. Cygorek}
\affiliation{Heriot-Watt University, Edinburgh EH14 4AS, United Kingdom}
\author{A. Vagov}
\affiliation{Theoretische Physik III, Universit{\"a}t Bayreuth, 95440 Bayreuth, Germany}
\affiliation{ITMO University, St. Petersburg, 197101, Russia}
\author{D. E. Reiter}
\affiliation{Institut f\"ur Festk\"orpertheorie, Universit\"at M\"unster, 48149 M\"unster, Germany}
\author{V. M. Axt}
\affiliation{Theoretische Physik III, Universit{\"a}t Bayreuth, 95440 Bayreuth, Germany}

\begin{abstract}
A Schr\"odinger-cat state is a coherent superposition of macroscopically distinguishable quantum states, in quantum optics usually realized as superposition of coherent states. Protocols to prepare photonic cats have been presented for atomic systems. Here, we investigate in what manner and how well the preparation protocols can be transferred to a solid state platform, namely a semiconductor quantum-dot--cavity system. In quantum-dot--cavity systems there are many disruptive influences like cavity losses, the radiative decay of the quantum dot, and the coupling to longitudinal acoustic phonons. We show that for one of the protocols these influences kill the quantum coherence between the states forming the cat, while for a second protocol a parameter regime can be identified where the essential characteristics of Schr\"odinger-cat states survive the environmental influences under conditions that can be realized with current equipment.
\end{abstract}

\maketitle

\section{Introduction}
\label{sec:Introduction}


Schr\"odinger cats are probably the most popular example of highly nonclassical, purely quantum mechanical states.
Realizing a coherent superposition of two macroscopically distinct states, where in analogy to Schr\"odinger's Gedankenexperiment \cite{Schroedinger1935} the cat is simultaneously dead and alive, remains a challenge due to the inevitable decoherence induced by the environmental coupling.
A general Schrödinger-cat state with two macroscopically distinguishable states $\vert\t{alive}\>$ and $\vert\t{dead}\>$ can be written as
\begin{align}
\label{eq:very_general_cat}
\vert\t{cat}\>=\mathcal{N}(\vert\t{alive}\> + e^{i\varphi}\vert\t{dead}\>)
\end{align}
with normalization $\mathcal{N}$ and phase $\varphi$.
These states find numerous applications in advanced quantum metrology \cite{Joo2011,facon2016sensitive}, quantum teleportation \cite{Enk2001}, quantum computation \cite{Ralph2003,Lund2008,Mirrahimi2014}, and quantum error correction algorithms \cite{Ofek2016}. Cat states being a superposition of more than two states \cite{leibfried2005creation,gao2010experimental,Sedov2020}, as well as phononic cat states \cite{Reiter2011} have been investigated. Schr\"odinger-cat states are a suitable platform to study the decoherence between two superposed quantum objects, in other words, to observe the quantum-to-classical transition \cite{Brune1996}. 
Therefore, Schrödinger-cat states are of fundamental interest in understanding the very foundations of quantum mechanics.

Schr\"odinger-cat states have to be sharply distinguished from incoherently superposed macroscopically distinct states, which are described by the density matrix
\begin{align}
\label{eq:mixture}
\r_{\t{mixture}}=\frac{1}{2}(\vert\t{alive}\>\<\t{alive}\vert+\vert\t{dead}\>\<\t{dead}\vert)\, ,
\end{align}
where the interference terms are missing.
This discrimination is best visible in the Wigner function. While in the cat state there are negative parts, which imply nonclassicality, the Wigner function of the classical incoherent mixture $\r_{\t{mixture}}$ is strictly positive.

\begin{figure}[t]
	\centering
	\includegraphics[width=0.45\textwidth]{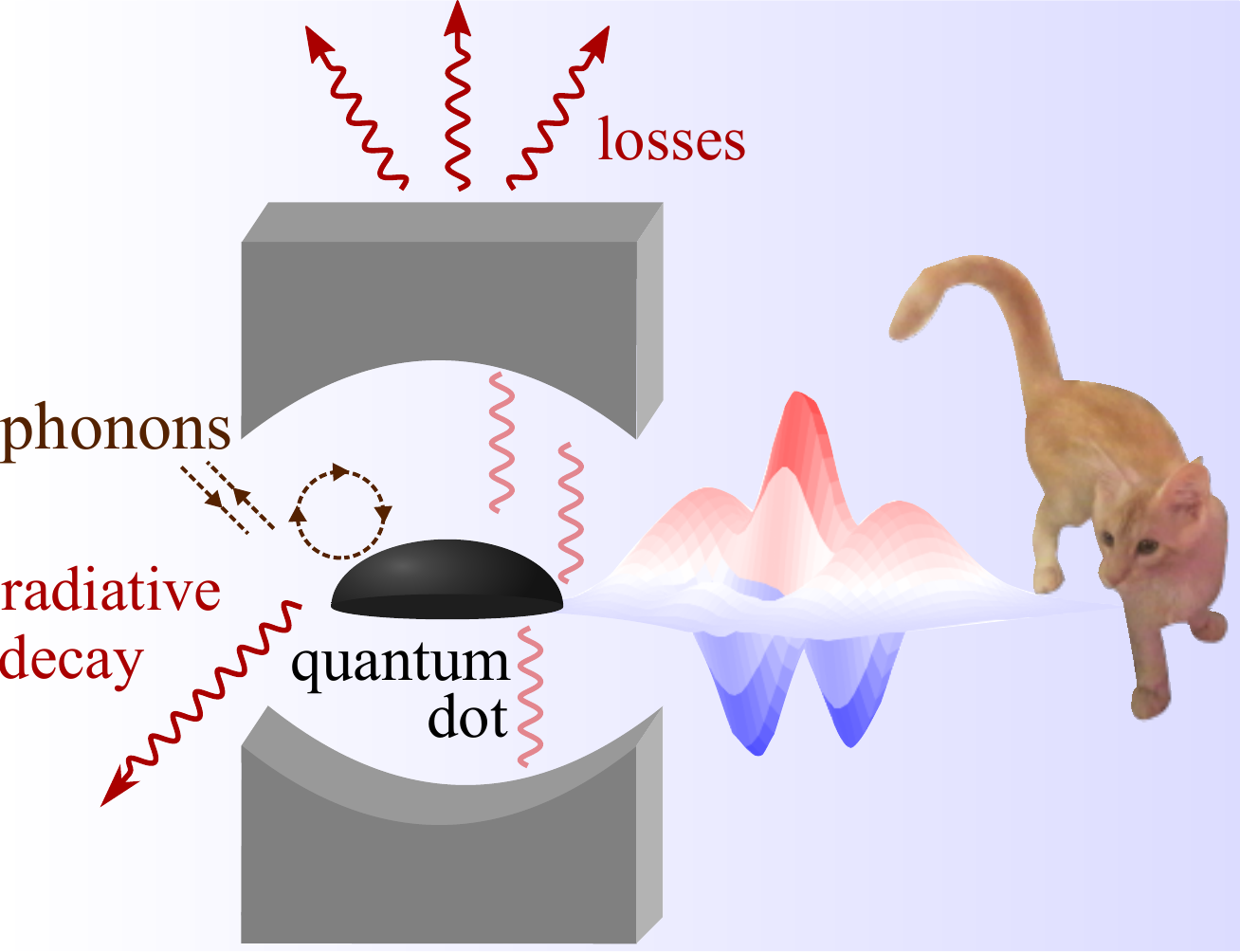}
	\caption{A Schrödinger cat state appearing in a QDC system. Photons are created by recombination of the QD exciton. The photoemission is controlled such that cats are created on demand. }
	\label{fig:sketch}
\end{figure}

Because of the fundamental and technological importance of Schrödinger cat states, their preparation has long been a research target.
Earlier efforts in this direction focused mostly on atom-based systems, where atoms are placed in an optical cavity \cite{Gerry2004,Law1996,Gea-Banacloche1990}. Recently, many concepts of atom quantum optics have been transferred to solid state systems, in particular to semiconductor quantum-dot--cavity systems (QDC) giving rise to the field of semiconductor quantum electrodynamics. QDCs have been already shown to be highly integrable on-demand emitters of photons in nonclassical states including high-quality single photons \cite{Michler2000,Santori2001,Santori2002,He2013,Wei2014Det,Ding2016,Somaschi2016,Schweickert2018,Hanschke2018,Cosacchi2019} and entangled photon pairs \cite{Akopian2006,Stevenson2006a,Hafenbrak2007,Dousse2010,Schumacher2012,
delvalle2013dis,Mueller2014,Munoz2015,Heinze2017,Orieux2017,Seidelmann2019}. QDC-based protocols for generating higher-order Fock states have also been developed \cite{Cosacchi2020b}.
An easy to use solid state-based source of Schrödinger cats would be a highly attractive extension of these achievements.

As atomic systems, QDCs have loss channels, namely the energetic loss channels provided by the radiative decay of a quantum dot (QD) and the finite cavity quality factor.
The main difference between QDC and atomic systems is the presence of longitudinal acoustic phonons, which is a well-known source of decoherence even at cryogenic temperatures of $T=4\,$K \cite{Ramsay2010a,Reiter2019}.
Hence, it is an open question whether also in the QDC platform the generation of Schr\"odinger-cat states is possible.

In this paper, we consider two protocols to prepare Schr\"odinger-cat states in a QDC. For both protocols we adapt existing preparation schemes and apply them to the QDC system. The first protocol is based on the proposal by Law and Eberly  \cite{Law1996}. To use this protocol in the QDC we solely rely on driving the QD by external laser pulses. To create a Schr\"odinger-cat state we need a precise timing of the arrival time of the pulses steering the QD-cavity coupling. We call the protocol DOD (\textbf{do}t \textbf{d}riven protocol). The second protocol is adapted from Gea-Banacloche \cite{Gea-Banacloche1990} and can be transferred to the QDC system by producing a coherent initial state. This can be achieved by driving the cavity, hence we call this protocol CAD (\textbf{ca}vity \textbf{d}riven protocol).

For both protocols, we analyze the impact of losses and phonon coupling in detail using realistic parameters that have been realized in current experiments. For the DOD we show that the losses are a detrimental influence on the sensitive coherence in a Schrödinger cat and the protocol can only produce mixed states [cf. Eq.~\eqref{eq:mixture}] under realsitic conditions. In contrast, for the CAD we find that even under realistic conditions it is possible to create a cat state and we identify experimentally accessible parameter regimes favorable for generating photonic Schr\"odinger cats.

Our work demonstrates that also in QDCs the preparation of Schr\"odinger-cat states is possible and we propose a protocol and specify a suitable parameter regime to prepare them. 

\section{Theoretical Model}
\label{sec:Model}


The QDC system can be well modeled by a driven two-level QD coupled to a single photon mode \cite{Reiter2019,Cosacchi2020b}. The corresponding Hamiltonian then reads
\begin{align}
\label{eq:2LS}
H=\,&H_{\t{QDC}}+H_{\t{driving}}+H_{\t{AC-Stark}}\, .
\end{align}
The QDC is described by the Jaynes-Cummings model
\begin{align}
H_{\t{QDC}}=\hbar\w_X \XX
+\hbar\w_C a^\dagger a+\hbar g\left(a\s_X^\dagger+a^\dagger\s_X\right)\, ,
\end{align}
where $\X$ is the exciton state at energy $\hbar\w_{X}$.
$\s_X:=\,\GX$ is the transition operator between the excited state $\X$ and the ground state $\G$.
The energy of $\G$ is set to zero.
$a$ ($a^\dagger$) denotes the photonic annihilation (creation) operator.
The cavity frequency is denoted by $\omega_C$ and its coupling to the QD by $g$.
We consider two different forms of driving Hamiltonians for the two protocols
\begin{align}
H_{\t{driving}}=\,&
\begin{cases}
-\frac{\hbar}{2}\left(f_{\t{p}}^*(t)\s_X+f_{\t{p}}(t)\s_X^\dagger\right) & \t{DOD}\\
-\frac{\hbar}{2}\left(f_{\t{p}}^*(t) a+f_{\t{p}}(t) a^\dagger\right) & \t{CAD}\, .
\end{cases}
\end{align}
Effectively decoupling the QD from the cavity can be achieved by an AC-Stark pulse driving the QD \cite{Unold2004,Cosacchi2020b}, described by
\begin{align}
H_{\t{AC-Stark}}=-\frac{\hbar}{2}\left(f_{\t{AC-Stark}}^*(t)\s_X+f_{\t{AC-Stark}}(t)\s_X^\dagger\right)\, .
\end{align}
The exciting and Stark laser pulses are represented by the functions $f_{\t{p}}(t)$ and $f_{\t{AC-Stark}}(t)$, which are specified in Appendix~\ref{app:hamiltonian}.

We also account for the coupling to longitudinal acoustic (LA) phonons \cite{Besombes2001,Borri2001,Krummheuer2002,
Axt2005,Reiter2019} (as detailed in Appendix~\ref{app:hamiltonian}) as well as the radiative decay of the QD exciton and cavity losses.
Whenever we consider phonon effects, the phonons are assumed to be initially in thermal equilibrium at a temperature of $T=4\,$K.
The corresponding Liouville equation is solved in a numerically complete manner by employing a path-integral formalism (for details see Refs.~\onlinecite{Vagov2011,Barth2016,Cygorek2017}).
The parameters used in the calculations are given in Appendix~\ref{app:parameters}.

\section{Dot driven protocol - DOD}
\label{sec:QD}

In the optical realm, coherent states $\vert\a\>$ are the most classical states.
A general coherent superposition of two coherent states of the form
\begin{align}
\label{eq:general_cat}
\mathcal{N}(\vert\a\> + e^{i\varphi}\vert-\a\>)
\end{align}
with normalization $\mathcal{N}$ and phase $\varphi$ is one of the most common realizations of Schrödinger-cat states in quantum optics \cite{Gerry2004}.
Hence, we choose this realization as the target state for the DOD and set $\a=\pi/2$ and $\varphi=0$.
This choice ensures that the corresponding coherent states are distinct, while their average photon number is low enough that we expect the influence of cavity losses to be limited.

To prepare this target state we adapt the protocol from Ref.~\onlinecite{Law1996}, which is proposed to create arbitrary photonic states in a single-mode microcavity. However, the originally proposed protocol does not account for any loss channels. The requirements are a driven Jaynes-Cummings model with controllable driving $f(t)$ and coupling $g(t)$ between the two-level system and the cavity. To transfer the proposal to QDCs, a few obstacles have to be overcome. While time-dependent driving of a QDC is possible by applying appropriate laser pulses, controlling the QD-cavity coupling time-dependently remains a challenge. In particular, the protocol in Ref.~\onlinecite{Law1996} relies on a stepwise switching between $f(t)$ and $g(t)$, i.e., one has to be off, whenever the other one is on.

Accordingly, the challenge of implementing this protocol in a QDC is twofold: (i) the magnitude of the QD-cavity coupling has to be varied and (ii) the QD and cavity are supposed to be decoupled during the time the driving is on.
In Ref.~\onlinecite{Law1996}, the time intervals $\tau$, when either $f(t)$ or $g(t)$ are on, are kept constant. Only the products $f_i\tau$ and $g_i\tau$ in the $i$th interval are relevant for the success of the protocol. Therefore, problem (i) can be solved by varying the time interval while keeping the coupling constant, i.e., we use $g\tau_i$. Concerning problem (ii), the decoupling suggested in Ref.~\onlinecite{Law1996} can in principle be realized by inducing suitable Stark shifts, which is, however, highly demanding experimentally. As demonstrated for protocols to prepare higher-order Fock states \cite{Cosacchi2020b}, it can be advantageous to avoid the decoupling provided the desired goal can be achieved by short enough pulses. Indeed, when the switching induced by the laser driving $f(t)$ takes place on time scales shorter than the dynamics induced by the cavity, the action of the latter cannot interfere noticeably with the switching, even though the QD and the cavity are coupled. Furthermore, on such fast time scales, the precise shape of the pulse becomes irrelevant (see, e.g., Appendix A 1 of Ref.~\onlinecite{Cosacchi2020b}). Therefore, a Gaussian pulse with the same area as the rectangular $f_i\tau$ can be used, which vastly reduces the experimental demand. Its full width at half maximum (FWHM) is chosen to be $100\,$fs.
Note that the corresponding spectral width of the pulse is quite large. Nonetheless, for typical energetic spacings to higher lying exciton states \cite{Vagov2004} the two-level approximation to the QD still holds well \cite{Cosacchi2020b}.

\begin{figure}[t]
	\centering
	\includegraphics[width=0.45\textwidth]{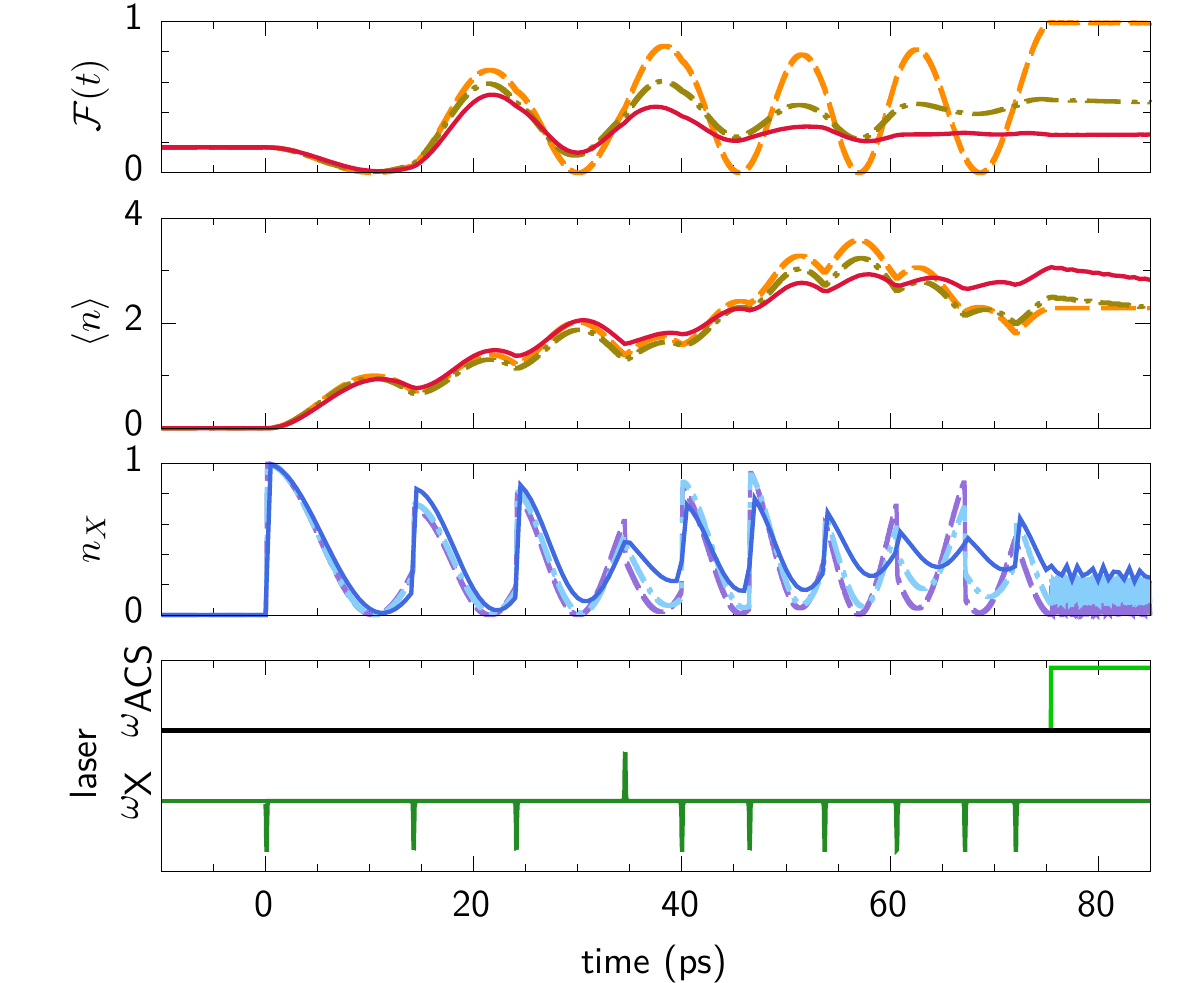}
	\caption{Dynamics of the QDC in the DOD. Panels from bottom to top: External pulses as well as the Stark pulse, 
	exciton occupation $n_X$, average photon number $\<n\>$ and time-dependent fidelity $\mathcal{F}(t)$ to the target Schr\"odinger-cat state in Eq.~\eqref{eq:general_cat}.
	Dashed lines: ideal case without phonons and losses. Dashed-dotted lines: without phonons but with losses. Solid lines: with phonons and losses.}
	\label{fig:LE_dyn}
\end{figure}

\subsection{The ideal case}
Figure~\ref{fig:LE_dyn} shows the dynamics of the QDC in the DOD. The lowest panel displays the sequence of laser pulses proposed to prepare the Schrödinger-cat state in Eq.~\eqref{eq:general_cat}. The pulse sequence is derived by solving the set of equations determining the protocol to prepare arbitrary states in Ref.~\onlinecite{Law1996}. After adapting the solution to pulses as explained above, one obtains the pulse areas and central peak times necessary to prepare the target state. In total, a series of ten $\pi$ pulses is applied. Note that relative phases of $\pi$ of the pulses are absorbed into the definition of the pulse areas and that the time difference between two subsequent pulses is the time $\tau_i$ where the cavity coupling $g$ takes effect. The arrival times and pulse areas are listed in Tab.~\ref{tab:LE}. After reaching the target state, the cavity needs to be decoupled from the QD in order to preserve the preparation, which is achieved by an AC-Stark pulse as shown in the lower panel of Fig.~\ref{fig:LE_dyn}.

\begin{table}[h]
\begin{center}
\caption{Pulse sequence for the DOD in Sec.~\ref{sec:QD}.
The times $t_c$ of their maxima and their pulse areas $\Theta$ are given.}
  \begin{tabular}{ l  c  r }
    \hline
    \hline	
	Number of the pulse & $t_c$ (ps) & $\Theta$ ($\pi$)\\
	\hline
	1   &    0.1     & -1\\                                                                                                   	2   &    14.2 & -1\\                                                                                                   	3   &    24.1 & -1\\                                                                                                   	4   &    34.6 & 1\\                                                                                                    	5   &    40.0 & -1\\                                                                                                   	6   &    46.5 & -1\\                                                                                                   	7   &    53.7 & -1\\                                                                                                   	8   &    60.7  & -1\\                                                                                                   	9   &    67.2 & -1\\                                                                                                   	10  &    72.1  & -1\\
    \hline
    \hline
  \end{tabular}
  \label{tab:LE}
\end{center}
\end{table}

The resulting time evolution of the exciton occupation $n_X$ is shown in blue in Fig.~\ref{fig:LE_dyn}, where the dashed lines correspond to the ideal case without phonons and losses. Each laser pulse partially excites the exciton, which then decays by photon creation. Accordingly, the average photon number $\langle n \rangle$, as shown in the second panel from top in Fig.~\ref{fig:LE_dyn}, increases after each pulse. 

To see whether we have created a Schr\"odinger-cat state, we consider the fidelity defined as 
\begin{equation}
	\mathcal{F}(\r_1,\r_2)=\left[\Tr\left(\sqrt{\sqrt{\r_1}\r_2\sqrt{\r_1}}\right)\right]^2
\end{equation}
for two arbitrary density matrices $\r_1$ and $\r_2$ \cite{Jozsa1994}. Setting one density matrix to the target Schrödinger-cat state in Eq.~\eqref{eq:general_cat}, we get a measure how close we are to this specific cat state. The results are shown in the topmost panel of Fig.~\ref{fig:LE_dyn}. We see that after the pulse sequence the fidelity reaches unity, implying that in the purely Hamiltonian ideal case without losses and phonons, this protocol is able to perfectly prepare the target state.

\begin{figure*}[t]
	\centering
	\includegraphics[width=\textwidth]{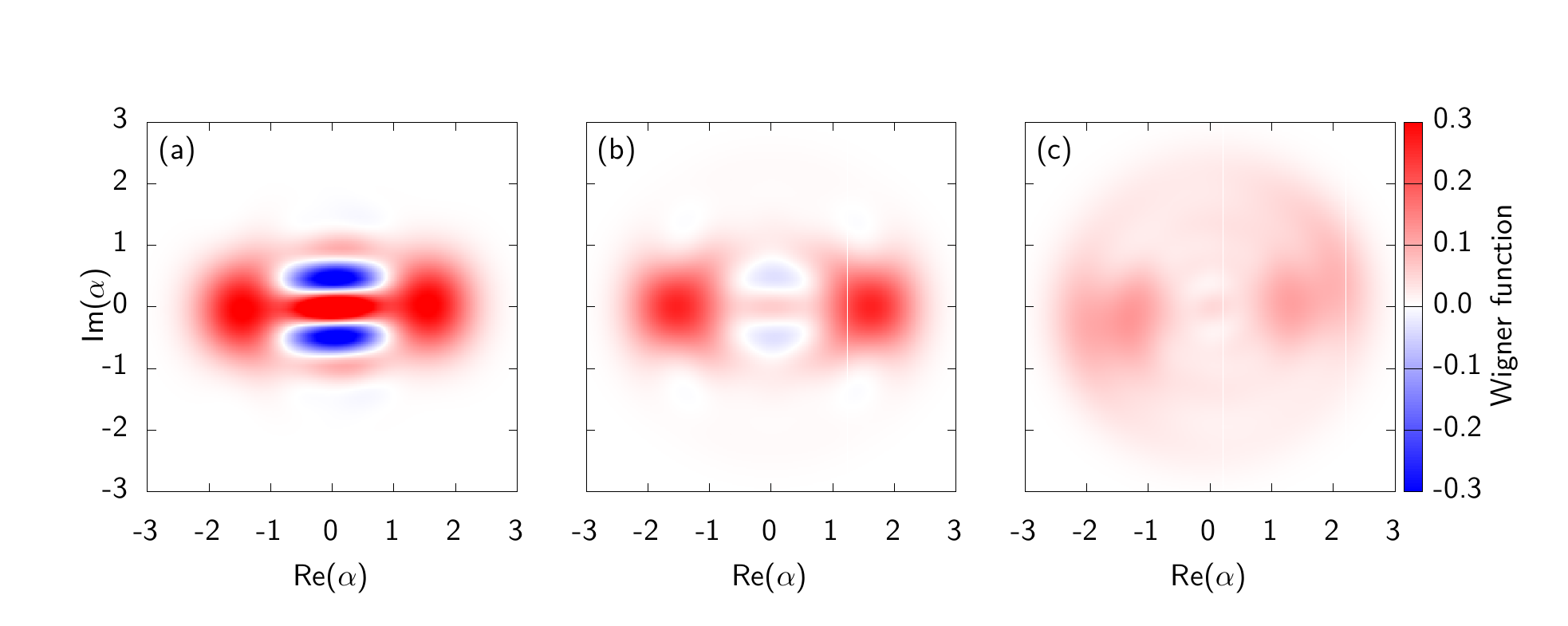}
	\caption{Wigner functions of the photonic states prepared by the DOD (a) for the ideal case, (b) including losses, and (c) taking both phonons and losses into account.
	The Wigner functions are calculated at the start time of the decoupling Stark pulse, i.e., at $t=75.5\,$ps.}
	\label{fig:LE_Wigner}
\end{figure*}

This is confirmed by looking at the Wigner function (definition in Appendix~\ref{app:Wigner}) in Fig.~\ref{fig:LE_Wigner} (a). We see all relevant features of a Schrödinger cat where the two Gaussians indicate the two macroscopically distinct states, here, coherent states, and the oscillations between them point to a coherent character of their superposition.

An important property of a cat state is its nonclassicality.
In general, the nonclassicality $\delta$ of a state can be measured by considering the negative part of its Wigner function $W(\alpha)$, since all Wigner functions corresponding to classical states are positive. The doubled volume of the integrated negative part of the Wigner function was introduced in Ref.~\onlinecite{Kenfack2004} as such a nonclassicality measure:
\begin{align}
\delta=\frac{\int(|W(\a)|-W(\a))d\a}{\int W(\a)d\a}\, . 
\end{align}
where $\delta=0$ implies a classical state. For the DOD we obtain $\delta=0.51$ for this Schrödinger cat, thus, indeed implying quantum features of the state. 

The results in Figs.~\ref{fig:LE_dyn} and \ref{fig:LE_Wigner} as well as the value of $\delta$ indicate that the adaption of the protocol in Ref.~\onlinecite{Law1996} to ideal QDCs with constant cavity coupling $g$ is accomplished successfully.

\subsection{Loss and phonon effects}

Next, we account for the loss channels. When taking the radiative decay of the QD and the cavity losses into account, the preparation fidelity drops to $\mathcal{F}=48.7\,\%$ (cf., dashed-dotted lines in Fig.~\ref{fig:LE_dyn}). Interestingly, the nonclassicality measure $\delta$ drops to $0.03$. Looking at the corresponding Wigner function in Fig.~\ref{fig:LE_Wigner} (b) reveals that the domains of negative values of the Wigner function have practically disappeared.

Phonons destroy even the remaining nonclassicality. While the fidelity in the case including all loss and phonon effects still yields $\mathcal{F}=25.0\,\%$ (see solid lines in Fig.~\ref{fig:LE_dyn}), $\delta$ is identically zero, thus indicating a classical state.
Indeed, the Wigner function in Fig.~\ref{fig:LE_Wigner} (c) shows two blurred macroscopically distinct states, here again coherent states, but no oscillations between them. Thus, they are superimposed incoherently and, therefore, closely resemble the statistical mixture in Eq.~\eqref{eq:mixture}.

The reason for this behavior lies in the nature of a Schrödinger-cat state, which involves the formation of a quantum mechanical superposition of distinct states with a well defined relative phase.
All processes diminishing this phase relation lead to a degradation of the cat state.
For typical QDCs, the cavity loss rate is much larger than the radiative decay rate, thus having greater impact on the preparation scheme.
Furthermore, the effective cavity loss rate is proportional to the photon occupation, thus degrading especially states with large multiphoton contribution, such as cat states.
The phonons then destroy any nonclassicality left after taking the other two processes into account.

From the analysis in this section it becomes clear that the fidelity alone is not sufficient to characterize a Schr\"odinger-cat state, since it may miss the essential feature of oscillations of the Wigner function to negative values. Therefore, it is necessary to simultaneously consider a nonclassicality measure like $\delta$.

In summary, the DOD is not suitable to prepare the Schrödinger-cat state in a QDC under realistic conditions, because the interference terms of a cat state do not survive the environmental coupling. Even at $T=4\,$K, phonon effects destroy the coherent superposition of the two macroscopically distinct states. Note that it is interesting from a fundamental point of view that the phonons provide sufficiently strong environmental coupling necessary to facilitate the quantum-to-classical transition, which in this case is a transition from a Schrödinger cat as in Eq.~\eqref{eq:general_cat} to an incoherent mixture as in Eq.~\eqref{eq:mixture}.

\section{Cavity driven protocol - CAD}
\label{sec:cavity}

\begin{figure}[t]
	\centering
	\includegraphics[width=0.45\textwidth]{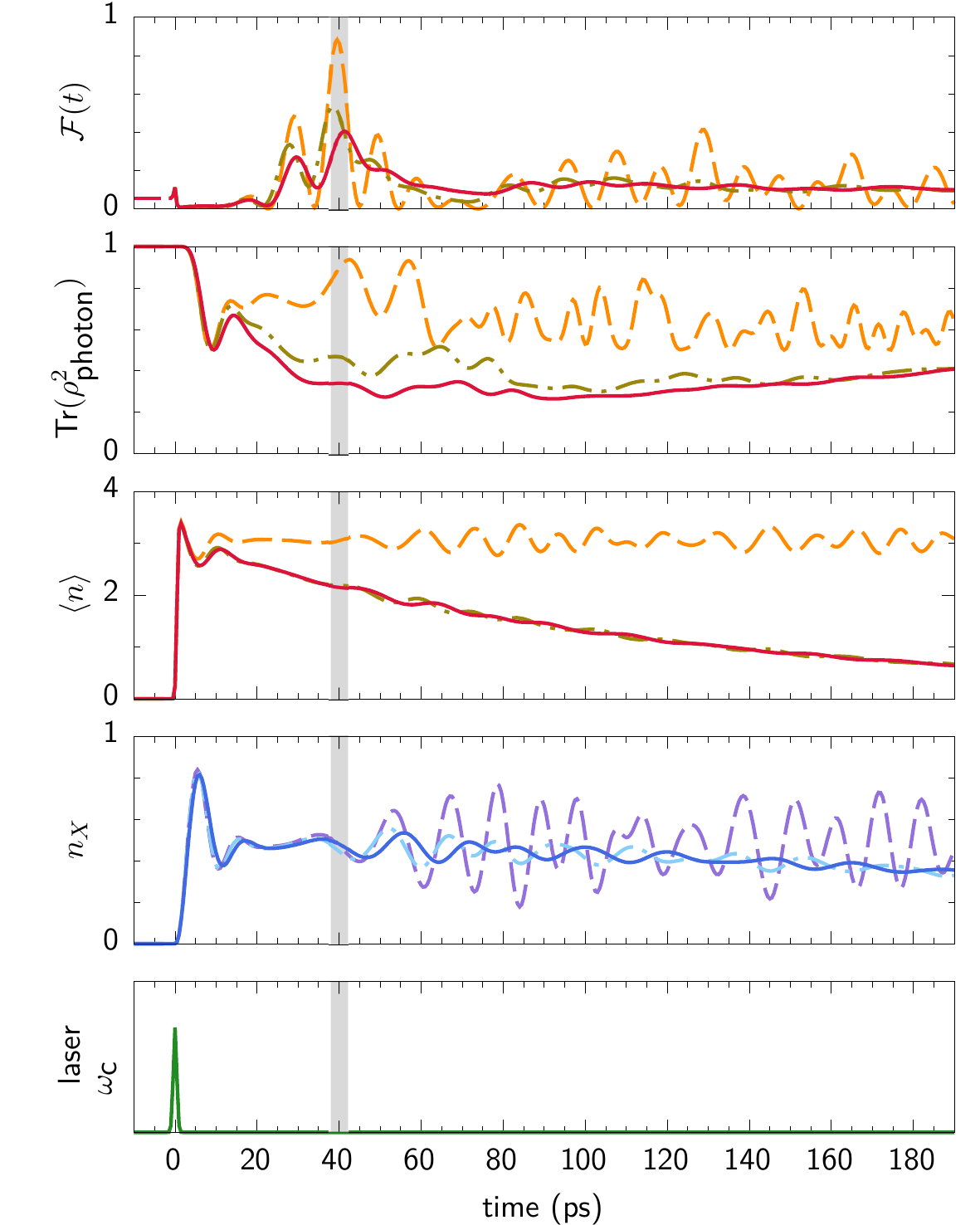}
	\caption{Dynamics of the QDC in the CAD.
	Panels from bottom to top: the exciting laser pulse, the occupation of the exciton $n_X$, the average photon number $\< n\>$, $\Tr(\r_{\t{photon}}^2)$ which indicates how close the photonic system is to a pure state, and the fidelity to the Schr\"odinger-cat state in Eq.~\eqref{eq:GB_cat}. Dashed lines: ideal case without phonons and losses. Dashed-dotted lines: without phonons but with losses. Solid lines: with phonons and losses. The temporal range where the fidelity reaches its maximum value in all three considered cases is shaded in gray.}
	\label{fig:GB_dyn}
\end{figure}

The CAD protocol is based on the ideas of Refs.~\onlinecite{Gea-Banacloche1990,Gea-Banacloche1991}, which showed that the textbook collapse-and-revival example in the Jaynes Cummings model has two additional very striking features: (i) at half the revival time the QD and photonic subspaces factorize and (ii) at precisely this time a Schrödinger-cat state appears in the cavity mode.

The main requirement for this cat state preparation scheme is a coherent state in the cavity mode as the initial state of the Jaynes-Cummings dynamics \cite{Gea-Banacloche1990}. The photonic state prepared at half the revival time is \cite{Gea-Banacloche1991}
\begin{align}
\label{eq:GB_cat}
\mathcal{N}\left(\vert\Phi_+\rangle + \vert\Phi_-\rangle\right)\, ,
\end{align}
which is another realization of a Schrödinger-cat state as defined in Eq.~\eqref{eq:very_general_cat}.
Here, $\mathcal{N}$ is a normalization constant and
\begin{align}
\label{eq:GB_states}
\vert\Phi_{\pm}\rangle=e^{-\frac{1}{2}\< n\>}\sum_{n=0}^\infty \frac{\< n\>^{n/2}}{\sqrt{n!}} e^{-i n \phi} e^{\mp i\pi \sqrt{\< n\> n}} \vert n\>
\end{align}
with $\< n\>=|\a|^2$ the average photon number of the initial coherent state and $\phi$ the phase of $\alpha$.
When using a real envelope Gaussian pulse as in our case, this phase is determined to be $3\pi/2$ as a short analytical calculation shows (cf., Appendix~\ref{app:phase}).
These two states are macroscopically distinguishable, but they are not coherent states as in Eq.~\eqref{eq:general_cat} because of the $\< n\>$- and $n$-dependent phase in Eq.~\eqref{eq:GB_states}. However, it should be noted that $\vert \Phi_\pm\>$ approaches in the limit $\< n\>\to\infty$ the coherent state $\vert \Phi_\pm\> \to e^{\mp i\< n\>\pi/2} \vert\mp i\a\>$ (cf., Ref.~\onlinecite{Gea-Banacloche1991}) and therefore the state in Eq.~\eqref{eq:GB_cat} becomes $\mathcal{N}(\vert i\a\> + e^{-i \< n\>\pi}\vert -i\a\>)$, which has the form given in Eq.~\eqref{eq:general_cat}.
The fidelity with which the state in Eq.~\eqref{eq:GB_cat} is reached in the Jaynes-Cummings dynamics at half the revival time is unity only in the limit $\< n\>\to\infty$ \cite{Gea-Banacloche1991}. Thus, in the ideal phonon and loss free case higher average photon numbers are favorable in this scheme. However, since the revival time is longer for higher mean photon numbers, the preparation time rises with increasing mean photon numbers. The system is thus exposed a longer time to losses before the end of the protocol. As a result of this trade-off situation, there is an optimal photon number when losses are accounted for.

\begin{figure*}[t]
	\centering
	\includegraphics[width=\textwidth]{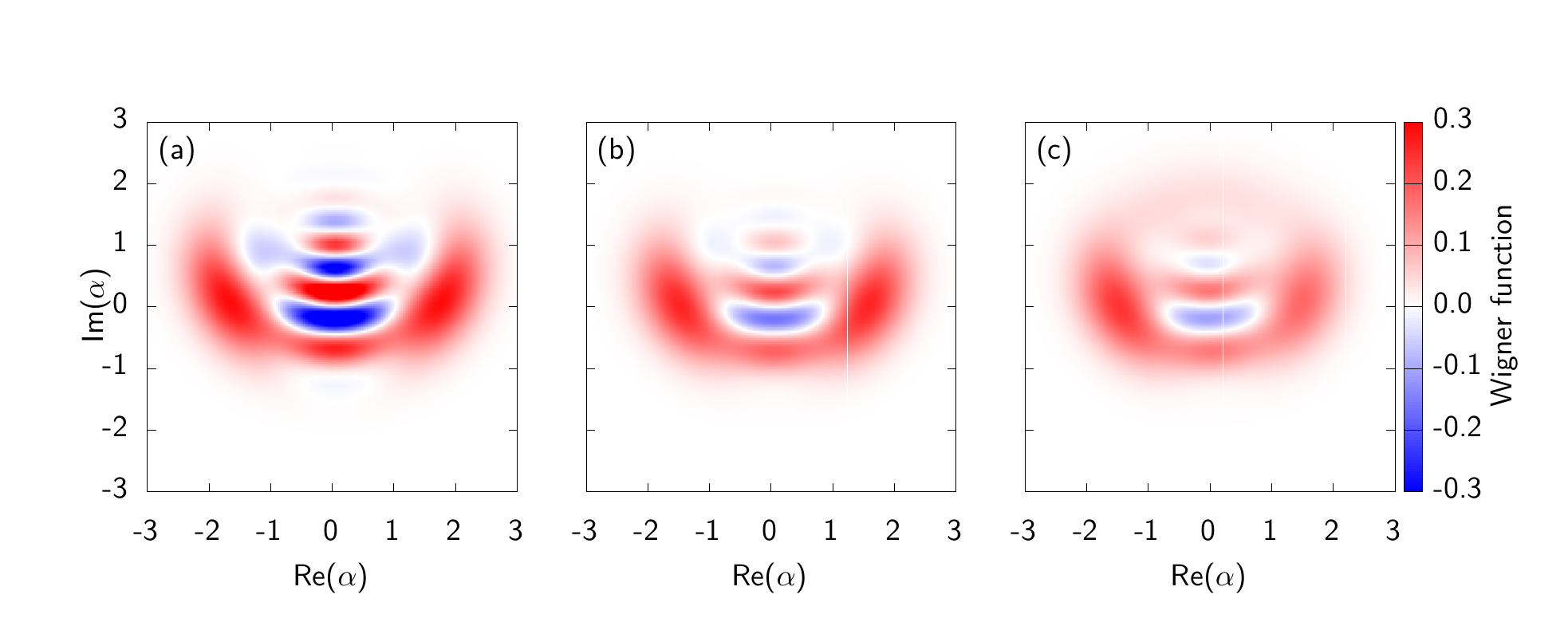}
	\caption{Wigner functions of the photonic states prepared by the CAD (a) for the ideal case, (b) including losses, and (c) taking both phonons and losses into account.
	The Wigner functions are calculated at the respective times of maximum fidelity (cf., Fig.~\ref{fig:GB_dyn}): $t=39.5\,$ps (ideal case), $t=38.5\,$ps (including losses), and $t=41.0\,$ps (including losses and phonons).}
	\label{fig:GB_Wigner}
\end{figure*}

Using this idea for a preparation protocol of Schrödinger-cat states in the microcavity mode, we have to face two tasks: (i) preparing a suitable initial state in the cavity and (ii) finding a suitable $\< n\>$ to achieve maximum fidelity to the cat state in Eq.~\eqref{eq:GB_cat}. Task (i) is solved easily, since it is textbook-knowledge \cite{Gerry2004} that driving an empty cavity with a classical laser yields a coherent state in the cavity. Now, the cavity in a QDC is not empty, but as long as the laser pulse driving the cavity mode is short enough compared with the dynamics induced by the coupling $g$, it may be approximated as such. We use a $1\,$ps-Gaussian pulse (cf., bottom panel of Fig.~\ref{fig:GB_dyn}) to drive the cavity mode and vary its pulse area $\Theta$ to analyze the success of the protocol depending on $\< n\>$ to tackle the second task (ii). Note that the average photon number of the prepared coherent state is connected to the laser pulse area via $\< n\>=|\a|^2=(\Theta/2)^2$.

\subsection{The ideal case}

Figure~\ref{fig:GB_dyn} shows the dynamics of the QDC in the CAD when driving the cavity with a pulse of area $1.2\pi$. As we show later this is the optimal pulse area when losses are taken into account. The dashed lines in Fig.~\ref{fig:GB_dyn} correspond to the ideal case.

For the chosen pulse area the average photon number, as shown in the middle panel in Fig.~\ref{fig:GB_dyn}, is about $3$. While this is not high enough to lead to a clear-cut collapse-and-revival signature, the exciton dynamics (blue curves) shows hints of this feature with a revival at about $80$~ps. The fidelity $\mathcal{F}$ shows oscillations within a bell-like envelope to reach its maximum of $\mathcal{F}=88.1\,\%$ at $39.5\,$ps after the pulse. At the same time, the photonic subsystem is close to a pure state as indicated by the near-unity value of the trace of the squared photonic density matrix.
Thus, the QD and photon subspaces factorize. To preserve the Schrödinger-cat state as it appears at this point in time, an additional QD-driving pulse is needed to effectively decouple the QD from the cavity. This is in analogy to the AC-Stark pulse shown for the DOD in Sec.~\ref{sec:QD}. 

Figure~\ref{fig:GB_Wigner} (a) shows the corresponding Wigner function evaluated at the time of maximum fidelity when the decoupling is evoked by a Stark pulse for the ideal case. Two macroscopically distinct states are clearly visible as elongated Gaussians corresponding to the states $\vert\Phi_\pm\rangle$ in Eq.~\eqref{eq:GB_states}. Oscillations to negative values between these two structures indicate a coherent superposition. Therefore, this state is clearly a Schrödinger-cat state.

As the CAD depends sensitively on the pulse areas, we plot in Fig.~\ref{fig:area_fid_noncl} (b) the maximum fidelity to the Schr\"odinger-cat state in Eq.~\eqref{eq:GB_cat} during the time evolution after the cavity preparation pulse as a function of the pulse area, i.e., the average photon number of the initial state. Having seen that the fidelity is not sufficient as a measure for obtaining a nonclassical photon state as the cat state, we additionally show in Fig.~\ref{fig:area_fid_noncl} (a) the nonclassicality measure $\delta$ at the time of maximum fidelity. 

\emph{\begin{figure}[t]
	\centering
	\includegraphics[width=0.45\textwidth]{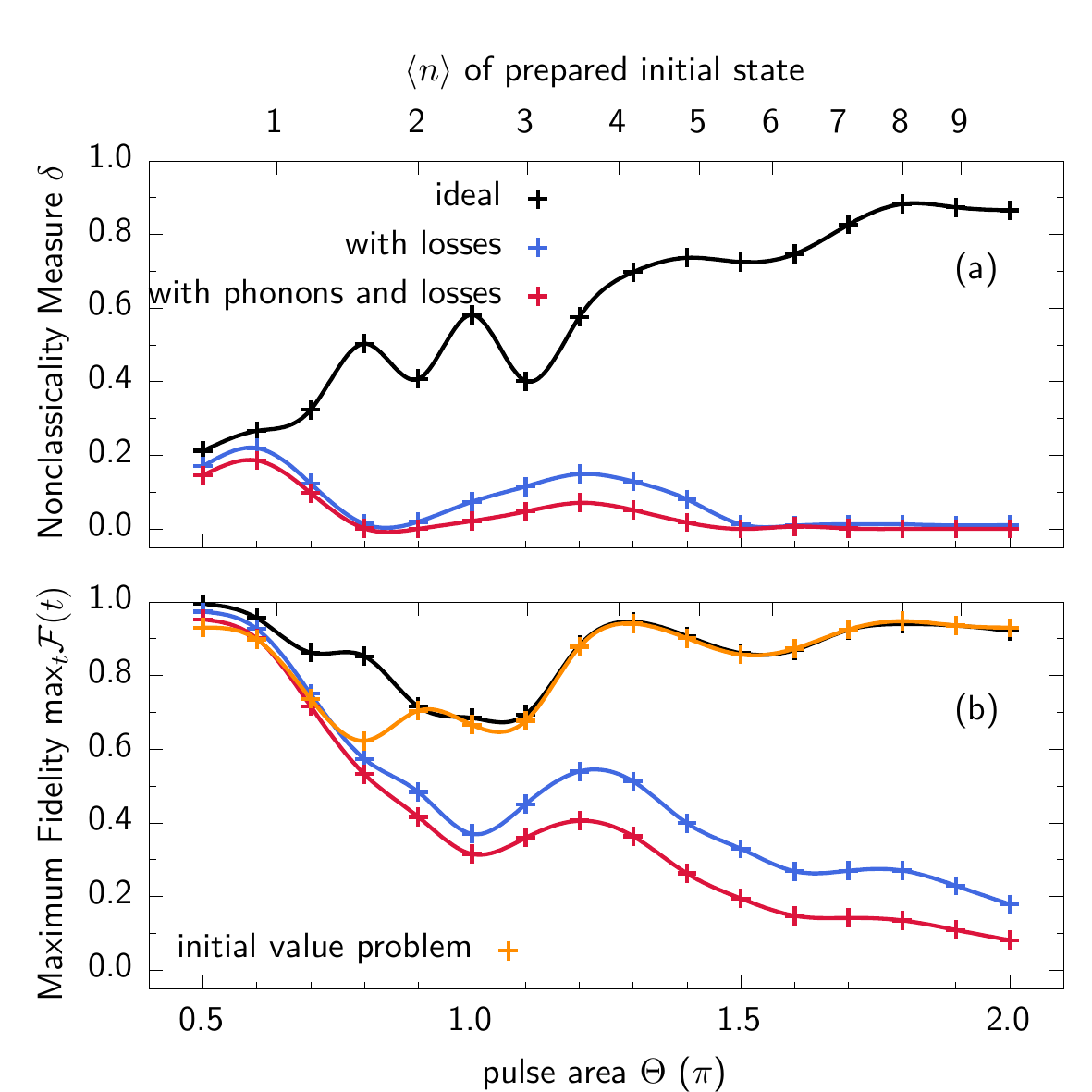}
	\caption{(a) Nonclassicality measure $\delta$ at the time of maximum fidelity and (b) maximum fidelity during the time evolution as a function of the pulse area, i.e., the average photon number of the initial state. Black: ideal case, blue: including losses but without phonons, red: with losses and phonons, orange: solution of the initial value problem .
	}
	\label{fig:area_fid_noncl}
\end{figure}}

For the ideal case (black lines in Fig.~\ref{fig:area_fid_noncl}) the nonmonotonic behavior of the fidelity is prominent. In contrast to the naive expectation that it rises monotonically with $\Theta$, i.e., $\< n\>$, it increases for $\< n \>\to 0$ and shows an oscillatory behavior for higher $\< n\>$. For decreasing $\< n\>$, the prepared state contains a larger contribution of the vacuum, while the target state in Eq.~\eqref{eq:GB_cat} itself shows more vacuum characteristics. Therefore, the fidelity approaches unity, while the nonclassicality measure $\delta$ decreases [cf., Fig.~\ref{fig:area_fid_noncl}(a)]. Thus, the state for low $\< n\>$ cannot be considered to be a genuine Schrödinger-cat, since the target state in Eq.~\eqref{eq:GB_cat} ceases to be a superposition of two macroscopically distinct states.

The nonmonotonic behavior of the maximum fidelity in Fig.~\ref{fig:area_fid_noncl}(b) starting at $\Theta=\pi$ has its origin in the dynamics of the fidelity as shown exemplary in Fig.~\ref{fig:GB_dyn}. The oscillation frequency within the bell-like envelope increases with rising $\Theta$ (not shown in the figure), while the maximum of the oscillation need not coincide with the maximum of the envelope. Therefore, $\t{max}_t\mathcal{F}(t)$ does not rise monotonically with $\Theta$, contrary to the naive expectation.

The proposal in Ref.~\onlinecite{Gea-Banacloche1990} is based on the assumption of a coherent state in the cavity mode as an initial state of the dynamics. The orange line in Fig.~\ref{fig:area_fid_noncl}(b) shows the fidelity corresponding to the solution of the initial value problem posed in Ref.~\cite{Gea-Banacloche1990}, i.e., without first preparing the initial state with a laser pulse. If such a coherent state is to be prepared in a cavity mode, an external laser pulse is necessary [cf., task (i) in Sec.~\ref{sec:cavity}]. For pulse areas greater than $\pi$, the maximum fidelity obtained after solving the initial value problem perfectly coincides with the result obtained in the ideal case of the preparation. This implies that indeed a coherent state is prepared in the cavity mode by the external driving. The deviations seen at smaller pulse areas have their origin in the finite length of the preparation pulse.

The overall rise of the nonclassicality measure $\delta$ with $\< n\>$, on the other hand, and its oscillations [cf., black line in Fig.~\ref{fig:area_fid_noncl}(a)] are a known feature \cite{Kenfack2004}. In particular, the oscillation is a signature of a nonzero phase due to a finite momentum of the cat state in $(q,p)$-representation that distinguishes 'standing' and 'moving' cats \cite{Kenfack2004}. Note that the Wigner function in $(q,p)$- and $(\t{Re}(\a),\t{Im}(\a))$-representation are connected by a factor of $2\pi\hbar$ \cite{Louisell1973,Cahill1969a,Cahill1969b}. Since the nonclassicality measure $\delta$ is a ratio of volumes of the Wigner function, it is independent of the representation.

\subsection{Loss and phonon effects}

The influence of cavity losses and radiative decay on the preparation using CAD is very strong. The dynamics of all photonic variables shown in Fig.~\ref{fig:GB_dyn} are damped (orange dashed-dotted line). The effect is even more pronounced when looking at the protocol as a function of $\< n\>$ in Fig.~\ref{fig:area_fid_noncl} (blue lines). In particular, at high $\< n\>$, where the highest fidelity in the ideal case is achieved, the losses have the greatest impact and the fidelity drops to almost zero. This is due to the fact that the effective loss rate for a Fock state $\vert n\>$ is proportional to the photon number $n$, i.e., $n\kappa$. Likewise, the nonclassicality measure $\delta$ becomes identically zero when the pulse area exceeds $1.5\pi$. Thus, in stark contrast to the ideal case, the limit $\< n\>\to\infty$ yields no Schrödinger cat at all.

Considering phonon effects on top of the loss influence further smoothens out the dynamics (cf., solid lines in Fig.~\ref{fig:GB_dyn}) and lowers both the fidelity and the nonclassicality of the target state [cf., red lines in Figs.~\ref{fig:area_fid_noncl}], while showing qualitatively the same behavior as in the case with losses but without phonons. These findings depend on the considered temperature, the GaAs material parameters, and the QD geometry and might differ for other parameter sets. In the case considered in this work, however, the loss effects have the most detrimental influence on the preparation of the target Schrödinger-cat state.

\begin{figure}[t]
	\centering
	\includegraphics[width=0.45\textwidth]{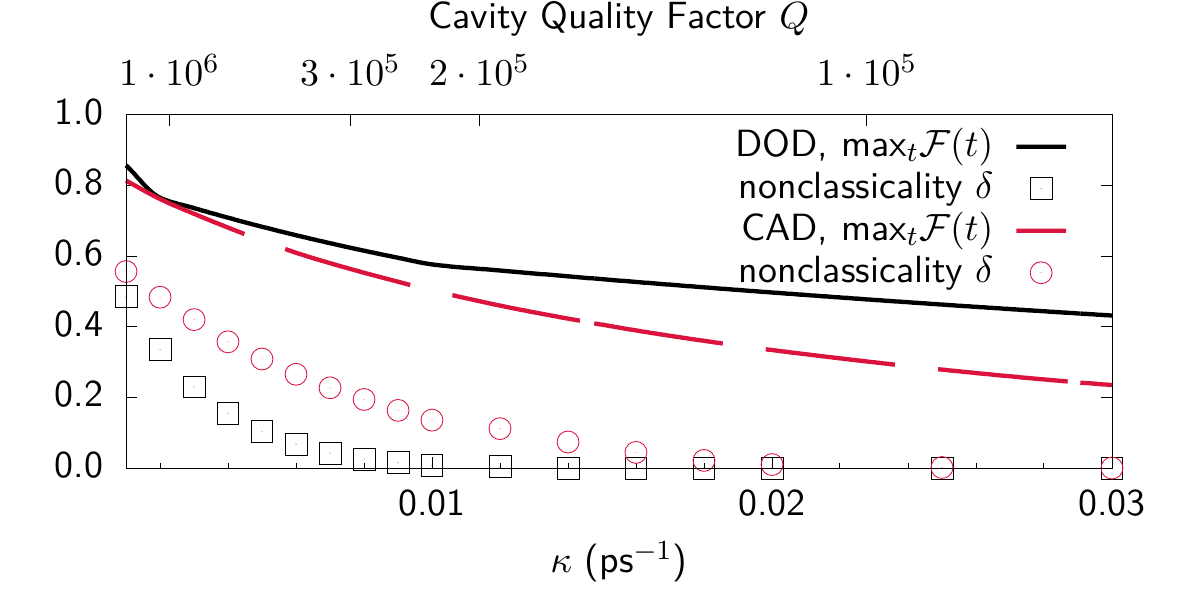}
	\caption{The maximum fidelity to the Schrödinger-cat states in Eqs.~\eqref{eq:general_cat} and \eqref{eq:GB_cat} in the case of the DOD (black solid line) and the CAD (red dashed line), respectively, and the nonclassicality measure $\delta$ (open rectangles and circles) as a function of the cavity loss rate $\kappa$.
	The cavity quality factor assuming a mode frequency of $\hbar\w_{\t{C}}=1.5\,$meV is displayed as a second axis.}
	\label{fig:kappa}
\end{figure}

This is further underscored by considering the dependence of the protocols' success on the cavity loss rate, as shown in Fig.~\ref{fig:kappa} for the case without phonons. For both the DOD and the CAD, the preparation fidelity rises monotonically with the quality factor, as does the nonclassicality measure $\delta$. The latter rises faster for the case of controlling the cavity, implying a comparatively higher robustness of this scheme with respect to cavity losses.
This is due to the fact that the total length of the DOD of about $~80\,$ps is roughly twice the length of the CAD of about $~40\,$ps.
Therefore, loss processes have less time to take effect in the CAD.
Furthermore, the DOD presumes ten operations, i.e., pulses which need to be timed exactly, whereas the CAD relies only on one pulse.
This also makes the CAD more stable against environmental influences and more attractive for experiments.

Note that it has been shown experimentally that a system with a Purcell factor roughly three orders of magnitude larger than in our case is well suited for the preparation of exotic photonic states.
Hofheinz \textit{et. al.} showed \cite{Hofheinz2009} that a so-called Voodoo cat state, a superposition of three coherent states, can be prepared in a system consisting of a superconducting qubit in a microwave resonator with a fidelity of $83\,\%$.
While such a setup has drawbacks, such as the wavelength of the emitted photonic state and the temperature in the mK-regime needed for the qubit to operate, this amazing result underscores the necessity of a high-quality resonator, both concerning the quality factor and the Purcell enhancement for the preparation of Schrödinger-cat or even more complicated Voodoo cat states.

Strikingly, in our QD-cavity system there is still a window of pulse areas, where the fidelity and the nonclassicality measure are rather high. The optimum pulse area yielding the maximum preparation fidelity under loss and phonon  influence for our parameters is at $\Theta=1.2\pi$, where both the fidelity and the nonclassicality measure have a maximum. This is due to a competition between the rising fidelity for $\< n\>\to\infty$ as predicted in Ref.~\onlinecite{Gea-Banacloche1991} and the inclusion of losses, which also become stronger for increasing $\< n\>$. As a check, whether we indeed have created a cat state at $\Theta=1.2\pi$, we have a look at the Wigner function in Fig.~\ref{fig:GB_Wigner} (c). The Wigner function clearly shows two macroscopically distinct states and oscillations between them, thus indicating a Schrödinger-cat state even when losses and phonons are accounted for under realistic conditions.

\section{Conclusion}
\label{sec:Conclusion}

We have investigated two protocols for the preparation of photonic Schrödinger-cat states in the light field mode of a quantum-dot--cavity system (QDC). While in atomic systems Schrödinger cats have been already prepared, we here adapted the protocols used in the atomic case to a solid state system. In the calculations, we considered realistic values for the cavity losses as have been reported in QDCs, which showed that the radiative decay and cavity losses can be quite detrimental to the preparation scheme. In contrast to atoms, in QDC devices also phonons play a role, which have great impact on the Schrödinger cat preparation. Therefore, a theoretical guidance on the feasibility to prepare cat states is of high importance. 

The first scheme relies on controlling the quantum dot with external laser pulses (DOD) by adapting Ref.~\cite{Law1996}. We developed a multi-pulse protocol for the QDC, where both the precise timing of the pulses and their mutual phases are of utmost importance. Most detrimental to the fidelity to the Schrödinger-cat state are cavity and radiative losses.
The environmental coupling to longitudinal acoustic phonons further reduces drastically the fidelity and completely destroys the coherence between the two states. Only an incoherent mixture of the two macroscopically distinct states with zero nonclassicality remains, such that this scheme is not suitable to prepare Schrödinger cats in realistic QDCs. We mention that this is different for superconducting qubits in microwave cavities, where similar protocols have been successfully employed \cite{Hofheinz2009}, because the quality factor of microwave cavities relative to the coupling strength is higher than in QDCs.
A similar boost of the quality factor would be needed to enable a Schrödinger-cat preparation with this protocol also in QDCs.

The second protocol exploits the internal dynamics of the Jaynes-Cummings model, where a Schrödinger-cat state can  be found naturally in the time evolution of the system \cite{Gea-Banacloche1990,Gea-Banacloche1991}. Only one pulse driving the cavity is necessary to prepare a single coherent state in the field mode (CAD), which serves as an initial state for the subsequent Jaynes-Cummings dynamics. While this protocol in the ideal case works best for high pulse areas, also the losses increase in this case such that no preparation is possible. Again, the losses are the main cause of destroying the cat, while the phonon effects are less dramatic than in the first protocol. Remarkably, for intermediate pulse areas between $\pi$ and $1.5\pi$, the coherences as well as the nonclassicality of the Schr\"odinger cats survive even under the influence of both losses and phonons.
Also in the CAD, a boost of the cavity quality factor would improve the characteristics of the prepared cat state.

Our results show that Schr\"odinger cats in QDCs can be prepared under realistic conditions with an easy to use protocol.

\acknowledgments
M. Co. thanks the cat Alfred for posing in Fig.~\ref{fig:sketch}.
This work was funded by the Deutsche Forschungsgemeinschaft (DFG, German Research Foundation) - project Nr. 419036043 and the University of Bayreuth in the funding program Open Access Publishing.

\appendix
\section{Coupling Hamiltonian}
\label{app:hamiltonian}
The external laser pulses are described by
\begin{align}
\label{eq:protocol}
f_{\t{p}}(t)=\,&\sum_m f_{m}^{\t{p}}(t-t_m) e^{-i\w_{\t{p}}(t-t_m)}\, .
\end{align}
$f_m^{\t{p}}(t)$ are the envelope functions of the pump fields and $\w_{\t{p}}$ the corresponding laser frequencies.
The product $\omega_\textrm{P} t_m$ is chosen as an integer multiple of $2\pi$.
The pump fields are Gaussian pulses with area $\Theta_j$
\begin{align}
f_j^{\t{p}}(t)=\frac{\Theta_j}{\sqrt{2\pi}\s}e^{-\frac{t^2}{2\s^2}}\,.
\end{align}
where $\s$ denotes the standard deviation.
It is connected to the full width at half maximum (FWHM) by FWHM$=2\sqrt{2\ln{2}}\s$.
Throughout this work, $\w_{\t{p}}=\w_{X}$ is assumed.
The AC-Stark pulses are assumed to be of rectangular shape
\begin{align} \label{eq:Stark}
&f_{\t{AC-Stark}}(t)=\,e^{-i\w_{\t{ACS}}t
}\nn
&\times\begin{cases}
0 & t<-\frac{\tau_{\t{length}}}{2} \\
f_{s} & -\frac{\tau_{\t{length}}}{2}\leq t \leq \frac{\tau_{\t{length}}}{2} \\
0 & t>\frac{\tau_{\t{length}}}{2}\, ,
\end{cases}
\end{align}
where $f_{s}$ denotes the field strength, i.e., the plateau height of the rectangular pulse, and $\tau_{\t{length}}$ its duration.
%

The AC-Stark pulses are tuned below the exciton line by $\w_{\t{ACS,X}}:=\w_{\t{ACS}}-\w_{\t{X}}$ that is chosen within the range of validity of the RWA.
The resulting shift of the exciton line can be calculated from the energies of the laser dressed states.

The QD is coupled to LA phonons  \cite{Besombes2001,Borri2001,Krummheuer2002,Axt2005,Reiter2014,Reiter2019}. 
\begin{align}
H_{\t{Ph}}=\sum_\q\hbar\w_\q b_\q^\dagger b_\q
+\sum_{\q}\left(\g_\q b_\q^\dagger+\g_\q^* b_\q\right)\XX\, ,
\end{align}
$b_\q^\dagger$ and $b_\q$ are the (bulk) phonon operators with wave vector $\q$ and energy $\hbar\w_\q$.
The deformation potential-type coupling to the electronic state is denoted by $\g_\q$.
This Hamiltonian is of the so-called pure dephasing-type \cite{Mukamel1995,Krummheuer2002}.
Many well-known phenomena emerge from the interaction described by this Hamiltonian, e.g., the phonon sideband in the QD emission spectrum \cite{Besombes2001,McCutcheon2016}, the damping of the Rabi oscillations \cite{Machnikowski2004,Ramsay2010a} as well as the renormalization of their frequency \cite{Kruegel2005,Ramsay2010b}.
Since our treatment of this Hamiltonian is numerically complete, all of these phenomena are included in our results.

Finally, we take radiative recombination of the excitons with rate $\gamma$ and cavity loss processes with rate $\kappa$ into account by introducing Markovian Lindblad-type operators
\begin{align}
\mathcal{L}_{O,\Gamma}\bullet=\Gamma\left(O\bullet O\+ -\frac{1}{2}\left\lbrace\bullet,O\+ O\right\rbrace_+\right)\, ,
\end{align}
where $\{\cdot,\cdot\}_+$ denotes the anti-commutator.
$O$ is a system operator and $\Gamma$ the decay rate of the associated loss process.

The full Hamiltonian then reads as
\begin{align}
\label{eq:H_full}
	H_{\t{full}}= H + H_{\text{Ph}}
\end{align}
with the system Hamiltonian $H$ as defined in Sec.~\ref{sec:Model}.
The dynamics of these systems are then described by the Liouville-von Neumann equation
\begin{align}
  \label{eq:Liouville-von Neumann}
  \frac{\partial}{\partial t} \r =\,&
-\frac{i}{\hbar}\{H_{\t{full}},\r\}_-
+\L_{a,\kappa}\r + \L_{\s_X,\g}\r\, ,
\end{align}
where $\{\cdot,\cdot\}_-$ denotes the commutator.

A path-integral formalism \cite{Makri1995a,Makri1995b,Vagov2011,Barth2016} is used to solve Eq.~\eqref{eq:Liouville-von Neumann} in a numerically complete manner.
Tracing out the phonon degrees of freedom analytically yields a phonon induced memory kernel for the subsystem of interest $H$ in Eq.~\eqref{eq:H_full}.
We call a solution ''numerically complete'' if a finer time discretization and considering a longer memory do not change the result noticeably.
Since the states considered in this paper are product states of the QD and number states of the cavity mode and therefore quite numerous, no solution within the path-integral framework could be obtained without the advances presented in Ref.~\onlinecite{Cygorek2017}.

%


\section{Definition of the optical Wigner function}
\label{app:Wigner}

The optical Wigner function, which is a function of the complex coherent amplitude $\a$, can be obtained as \cite{Louisell1973,Cahill1969a,Cahill1969b}
\begin{align}
W(\a)=2\t{Tr}\left[\r_{\t{photon}}D(\a)(-1)^{a^\dagger a}D(-\a)\right]\, ,
\end{align}
where $\r_{\t{photon}}$ is the photonic density matrix of the system and $D(\a)$ the coherent displacement operator.
The photonic density matrix is obtained by tracing out the phonon and QD degrees of freedom:
\begin{align}
\label{eq:rho_photon}
\r_{\t{photon}}=\t{Tr}_{\t{QD}}\left[\t{Tr}_{\t{Ph}}(\r)\right]\, .
\end{align}

Using the Fock basis and introducing $\r'_{\t{photon}}(\a):=D(-\a)\r_{\t{photon}}D(\a)$, this expression simplifies to
\begin{align}
W(\a)=2\sum_n[\r'_{\t{photon}}(\a)]_{nn}(-1)^n\, .
\end{align}


\section{Parameters}
\label{app:parameters}

For the numerical calculations we use typical parameters for self-assembled strongly confined GaAs/In(Ga)As QDs \cite{Krummheuer2005,Cygorek2017}.
Other relevant parameters are summarized in Tab.~\ref{tab:par}.
Assuming a mode frequency of $\hbar\w_{\t{C}}=1.5\,$eV, the cavity loss rate $\kappa$ corresponds to a quality factor $Q\approx268,000$, which has been reported in the experiments in Ref.~\onlinecite{Schneider2016}.
The phonons are assumed to be initially in thermal equilibrium at a temperature of $T=4\,$K, whenever phonon effects are considered in this work.

On a time scale of $\approx3\,$ps, the phonon induced memory kernel for GaAs/In(Ga)As QDs of $6\,$nm diameter at $T=4\,$K decays to zero \cite{Vagov2011,Barth2016,Cygorek2017}.
To obtain numerically complete converged results, a two-grid strategy is employed for the time discretization.
Details can be found in Appendix A 3 of Ref.~\onlinecite{Cosacchi2020b}.

\begin{table}[h]
\begin{center}
\caption{Relevant system parameters.}
  \begin{tabular}{ l  c  r }
  
    \hline
    \hline	
    QD-cavity coupling (meV) & $\hbar g$ & 0.1 \\
    Cavity loss rate (ps$^{-1}$) & $\kappa$ & 0.0085 \\
    QD radiative decay rate (ps$^{-1}$) & $\gamma$ & 0.001 \\
    \hline
    \hline
  \end{tabular}
  \label{tab:par}
\end{center}
\end{table}



\section{Calculation of the phase in the CAD}
\label{app:phase}

The phase $\phi$ in Eq.~\eqref{eq:GB_states} is the phase of the coherent state which is prepared in the cavity mode by the initial laser pulse driving the cavity.
Since the pulse used to prepare the coherent state is short compared with the scale of the dynamics induced by the coupling to the QD, we can neglect the latter in the analysis of the preparation.
In a frame co-rotating with the laser frequency, the Hamiltonian thus reduces to
\begin{align}
H_{\t{driving}}(t)=\,&-\frac{\hbar}{2}(f_{\t{p}}^*(t) a + f_{\t{p}}(t) a\+)\, .
\end{align}
Up to second order in the time increment $\Delta t$, the time evolution operator reads
\begin{align}
U(t+\Delta t,t) =&\, e^{-\frac{i}{\hbar}H_{\t{driving}}(t)\Delta t}\nn
=&D(u(t)\Delta t)
\end{align}
with $u(t):=\frac{i}{2}f_{\t{p}}(t)$ and the coherent displacement operator $D$.
Using the relation $D(\a)D(\b)=\exp{[\t{Im}(\a\b^*)]}D(\a+\b)$ and noting that the phase vanishes since $u(t)$ is purely imaginary for real pulse envelopes, one obtains for a pulse with center $t_c$ that is chosen such that at $0$ and $\tau_{\t{max}}$ the envelope is essentially zero (again up to second order in $\Delta t$)
\begin{align}
U(\tau_{\t{max}},0)=D(\a(\tau_{\t{max}}))
\end{align}
with
\begin{align}
\a(\tau_{\t{max}})=&\,\int_0^{\tau_{\t{max}}} u(t) dt\nn
=&\frac{i}{2}\int_0^{\tau_{\t{max}}} \frac{\Theta}{\sqrt{2\pi}\s}e^{-\frac{(t-t_c)^2}{2\s^2}} dt
\approx\frac{i}{2}\Theta
=\frac{\Theta}{2}e^{-i\phi}
\end{align}
with $\phi=3\pi/2$.

\bibliographystyle{appa}
\bibliography{bib,DorisCat}
\end{document}